\documentclass[aps,prl,twocolumn,superscriptaddress,floatfix,preprintnumbers]{revtex4}

\newcommand{\bea}{\begin{eqnarray}}
\newcommand{\eea}{\end{eqnarray}}
\newcommand{\be}{\begin{equation}}
\newcommand{\ee}{\end{equation}}
\newcommand{\mw}   {\mbox{$m_{W}$}}
\newcommand{\mwsq}   {\mbox{$m_{W}^2$}}

\newcommand{\mz}   {\mbox{$m_{Z}$}}
\newcommand{\mzsq}   {\mbox{$m_{Z}^2$}}

\newcommand{\muf}{\mu_F}
\newcommand{\mur}{\mu_R}

\newcommand{\oa}{\mbox{${\cal O}(\alpha)$}\,}
\newcommand{\oas}{\mbox{${\cal O}(\alpha_s)$}\,}
\newcommand{\oaas}{\mbox{${\cal O}(\alpha\alpha_s)$\,}}

\usepackage{amsmath}
\usepackage{graphicx}
\usepackage{xcolor}
\usepackage[normalem]{ulem}
\usepackage[utf8]{inputenc}

\begin{document}

\preprint{OUTP-20-07P, TIF-UNIMI-2020-17}

\title{NNLO QCD$\times$EW corrections to on-shell $Z$ production}

\author{Roberto Bonciani}
\email[]{roberto.bonciani@roma1.infn.it}
%\homepage[]{Your web page}
%\thanks{}
%\altaffiliation{}
\affiliation{Universit\`a di Roma ``La Sapienza'', P.le Aldo Moro 5, 00185 Roma, Italy}
\affiliation{INFN Sezione di Roma, P.le Aldo Moro 5, 00185 Roma, Italy}

\author{Federico Buccioni}
\email[]{federico.buccioni@physics.ox.ac.uk}
%\homepage[]{Your web page}
%\thanks{}
%\altaffiliation{}
\affiliation{Rudolf Peierls Centre for Theoretical Physics, 
Clarendon Laboratory, Parks Road,
Oxford OX1 3PU,
UK}

\author{Narayan Rana}
\email[]{narayan.rana@mi.infn.it}
%\homepage[]{Your web page}
%\thanks{}
%\altaffiliation{}
\affiliation{INFN Sezione di Milano, Via Celoria 16, 20133 Milano, Italy}

\author{Alessandro Vicini}
\email[]{alessandro.vicini@mi.infn.it}
%\homepage[]{Your web page}
%\thanks{}
%\altaffiliation{}
\affiliation{Dipartimento di Fisica ``Aldo Pontremoli'', University of Milano, Via Celoria 16, 20133 Milano, Italy}
\affiliation{INFN Sezione di Milano, Via Celoria 16, 20133 Milano, Italy}

\date{\today}

\begin{abstract}
We present the first analytical results for the ${\cal O}(\alpha\alpha_s)$ corrections to the total cross section 
for the inclusive production of an on-shell $Z$ boson at hadron colliders. 
We include the complete set of contributions, with photonic and massive weak gauge boson effects, 
which have been computed in analytical form and expressed in terms of polylogarithmic and elliptic integrals. 
We present numerical results, relevant for the precision studies at the LHC. 
These corrections increase the accuracy of the predictions and contribute to the reduction of the QCD component of the theoretical uncertainty.
\end{abstract}

%\keywords{}

\maketitle

%%%%%%%%%%%%%%%%%%%%%%%%%%%%%%%%%%%%%%%%%%%
The production in hadronic collisions of a pair of leptons, each with large transverse momentum, is known as Drell-Yan (DY) process 
and it plays a fundamental role for our understanding of Quantum Chromodynamics (QCD) as the theory of the strong interactions. 
The lepton pair acts as a probe of the initial-state proton structure: it allows for the measurement of the proton collinear parton density functions (PDFs) 
and for the study of the QCD dynamics from the analysis of the lepton-pair transverse momentum distribution. 
The kinematical distributions of the final-state leptons allow for precision tests of the electroweak (EW) Standard Model (SM), 
with the determination of the weak mixing angle and of the masses $m_{W,Z}$ and decay widths $\Gamma_{W,Z}$ of the $W$ and $Z$ bosons.

The production of an on-shell $Z$ boson represents a special kinematical configuration of the full neutral-current DY process. 
It plays a major phenomenological role because: $i)$ it provides insight into the $Z$ boson properties, $ii)$ it offers a constraint 
of the absolute normalization of the proton PDFs and, in turn, $iii)$ of the absolute calibration for the kinematical distributions, the tails of which might 
reveal potential signals of new Physics beyond the SM. 

This process is described in lowest order by quark-antiquark annihilation into a $Z$ boson via EW interaction. 
The evaluation of the next-to-leading order (NLO) \cite{Altarelli:1979ub}, 
next-to-next-to-leading order (NNLO) \cite{Hamberg:1990np,Harlander:2002wh}, and 
next-to-next-to-next-to-leading order (N$^3$LO) \cite{Ahmed:2014cla,Li:2014bfa} QCD corrections to 
the production of an on-shell gauge boson, supplemented by the resummation of the logarithmically enhanced terms 
due to soft gluon emission \cite{Sterman:1986aj,Catani:1989ne,Catani:1990rp,Moch:2005ky,Ravindran:2006cg,Catani:2014uta,H.:2020ecd}, 
allows for the accurate estimate of  
the total cross section, the reduction of the impact of QCD uncertainty and the precise assessment of its actual size. 
The best available result for the inclusive production of a virtual photon includes up to N$^3$LO QCD corrections \cite{Duhr:2020seh}. It shows a dependence on 
the QCD renormalization ($\mur$) and factorization ($\muf$) scale choices at the sub-percent level for virtualities $Q>70$ GeV and 
at the percent level for smaller $Q$ values, with a stronger sensitivity to the choice of the factorization scale. 

In this high-precision QCD framework, the inclusion of EW effects becomes mandatory. 
The NLO-EW corrections to the DY process have been computed in \cite{Baur:2001ze,CarloniCalame:2007cd,Arbuzov:2007db,Dittmaier:2009cr,Buonocore:2019puv} 
and are comparable in size to the NNLO-QCD effects. The theoretical uncertainty associated with missing higher-order EW corrections is formally 
at the NNLO-EW level and it is significantly reduced compared to the leading order (LO) case. Using different input parameters as a mean to estimate the size of 
missing higher-order EW effects, one finds that the LO variation is at the ${\cal O}(3.5\%)$ level, whereas the NLO-EW one is reduced down to the ${\cal O}(0.5\%)$ level.
The higher-order QCD predictions are only LO accurate from the point of view of the EW interaction,
thus they suffer from the uncertainty associated with different choices of
input parameters. If we consider the NNLO-QCD prediction, supplemented with the NLO-EW one, we find a scheme uncertainty at 
the ${\cal O}(0.88\%)$ level. 
This value is significant for any precision test and comparable to the residual QCD uncertainty.
On the other hand, a specular discussion applies to the NLO-EW corrections, which are only LO from the point of view of the strong interaction. 
They suffer from large uncertainties under variations of the factorization scale. The canonical $\muf$ variation by a factor 2 about its 
central value
yields a change of 
the LO cross section by $\pm 18\%$ and, in turn, a change of the NLO-EW correction at the ${\cal O}(0.5\%)$ level. 
In order to increase the control on the theoretical error, it is therefore mandatory to include in the analysis the mixed QCD-EW corrections, 
since they stabilize both the dependence on the QCD scales of the higher-order EW corrections and the dependence on the EW input parameters of 
the higher-order QCD corrections. 

The mixed corrections to the DY process in the resonance region have been studied
in the so-called pole approximation in Refs.\cite{Dittmaier:2014qza,Dittmaier:2015rxo},
where QCD and EW effects are factorized between production and decay of the vector boson.
As far as the production of an on-shell $Z$ boson is concerned,
the exact QCD-QED corrections have been considered
in Refs.\cite{deFlorian:2018wcj,Cieri:2018sfk,Delto:2019ewv},
while in \cite{Bonciani:2019nuy} we have computed the EW effects in the $q\bar{q}$ initiated channels.
The complete set of QCD$\times$EW effects have been presented in Ref.\cite{Buccioni:2020cfi} at the fully differential level,
using a combination of analytical and numerical techniques. 
As for the off-shell neutral-current DY, the QCD$\times$QED corrections to the production of a pair of neutrinos have been discussed in Ref.\cite{Cieri:2020ikq}.

In this letter we present the totally inclusive cross section for the production of a
single on-shell $Z$ boson in hadron-hadron collisions, including the \oaas mixed QCD-EW corrections
stemming from all the relevant partonic channels. The inclusion of the \oaas corrections increases the
accuracy of the prediction and it reduces the impact of the residual theoretical uncertainties.
The results have been fully computed in analytical form and expressed in terms of polylogarithms and elliptic integrals, 
requiring the evaluation of new two-loop Master Integrals (MI) that were not available in the literature.

%%%%%%%%%%%%%%%%%%%%%%%%%%%%%%%%%%%%%%%%%%%% 

\section{Theoretical framework}

%%%
At hadron colliders, the inclusive production cross section $\sigma_{tot}$ of an on-shell $Z$ boson $(pp\to Z+X)$ is written, based on the factorization theorem, as 
\begin{align}
&\sigma_{tot} (\tau) = \sum_{i,j\in q,\bar q, g, \gamma}
\int {\rm d}x_1 {\rm d}x_2 f_i (x_1) f_j (x_2) \sigma_{ij} (z) \, ,                 
\label{eq:sigmatot-bare}
\end{align}
where the sum runs over all possible initial state partons, namely quarks, gluons and photons. The ratios $\tau=\frac{m_Z^2}{S}$ and $z=\frac{m_Z^2}{\hat{s}}$ 
compare the $Z$ boson mass, $\mz$, with $S$ and $\hat{s}$, the hadronic and partonic center of mass energy squared, respectively. 
The variables $S$ and $\hat{s}$ are related by $\hat{s}=x_1 x_2 S$ through the Bjorken momentum fractions $x_1, x_2$.
The hadron-level result is obtained via the convolution of the cross section $\sigma_{ij}$ of the partonic process  $ij\to Z+X$ with 
the physical parton densities $f_i (x)$.
The cross sections are inclusive over the additional partons radiated present in the final state 
and collectively indicated as $X$. However, we do not include the processes with the emission of one extra massive on-shell gauge boson, 
as their measurement depends on the details of the experimental event selection. The process that contributes at the lowest perturbative 
order (Born approximation) is quark-antiquark annihilation, while additional channels open in higher perturbative orders. 
We include all the corrections of \oaas relative to the Born process. Each partonic cross section admits a double expansion in 
the electromagnetic and strong coupling constants, $\alpha$ and $\alpha_s$, respectively.
At NNLO-QCD$\times$EW we need to consider two-loop virtual corrections to the $q\bar q\to Z$ process (dubbed double-virtual), 
single virtual corrections to the processes with one additional parton in the final state (dubbed real-virtual), and the \oaas part
\footnote{
  We observe that the tree-level amplitude of some partonic processes is mediated by the exchange of both
  QCD and EW bosons, so that it simultaneously contributes to different perturbative orders.
  We isolate the \oaas terms.
  }
of the tree-level cross section of the processes with two additional real partons in the final state (dubbed double-real). 
The charged-current weak interaction couples up- and down-type fermions, so that it is natural to observe a change of flavour in the internal 
lines when a $W$ boson is exchanged, but also down type quarks in the final state of a process initiated by up-type quarks (and viceversa). 
Their impact eventually depends on the values of the $Z$ coupling to the different flavours, in such a way that their complete 
evaluation requires the inclusion of one complete quark doublet in the hadron-level cross section. 
In the present study we consider a diagonal CKM matrix and our predictions are given in the four-flavor scheme,  
with the first two fermionic families.

The prediction of the hadron-level cross section requires to express the bare couplings and masses
in terms of physical parameters via renormalization.
The choice of the background field gauge (BFG) \cite{Denner:1994xt}
allows to restore the validity of $U(1)_{em}$-like Ward identities
between the vertex corrections and the external quark wave function corrections in the full EW model.
We explicitly verify these Ward identities at \oaas.
Only the charge renormalization of the $Z$ boson couplings to quarks is then needed.
The ultraviolet (UV) renormalization of the calculation at \oaas
affects the weak charge of the LO $q\bar q\to Z$ amplitude.
The two processes $q\bar q\to Zg$ and $qg\to qZ$, involving an external gluon,
require the \oa renormalization of the $Z$ coupling to the quarks.
Instead, there is no \oas renormalization of this coupling
in the processes involving an external photon ($q\bar q\to Z\gamma$ and $q\gamma\to qZ$).
The charge renormalization counterterm in fact starts at \oa
and it explicitly depends on the choice of input parameters used to parameterize the physical Lagrangian.
The \oas terms are absent.
If we choose to express $(g,g',v)$,
the $SU(2)_L$ and $U(1)_Y$ couplings and the Higgs field vacuum expectation value,
in terms of $(G_\mu,\mw,\mz)$ (dubbed $G_\mu$-scheme), where $G_\mu$ is the Fermi constant,
then the weak charge renormalization is achieved by the replacement
\be
\frac{g_0}{c_0} Z_{ZZ}^{1/2}
\to
\sqrt{4\sqrt{2} G_\mu \mz^2}\,
\left(1-\frac12 \Delta r +\frac12 \delta g_Z\right)\, .
\label{eq:chargeGmu}
\ee
We denote with a $0$ subscript all the bare quantities.
We abbreviate with $c=\mw/\mz$ the cosinus of the weak mixing angle ($s^2=1-c^2$), and we define
$ \delta g_Z\equiv \delta Z_{ZZ}+\delta e^2/e^2+(s^2-c^2)/(c^2) (\delta s^2/s^2) $
where
$Z_{ZZ}=1+\delta Z_{ZZ}$ is the $ZZ$ wave function renormalization constant,
$\Delta r$ is a finite correction \cite{Sirlin:1980nh}
expressing the relation between the Fermi constant and the muon decay amplitude,
$\delta s^2=c^2\left(\delta\mz^2/\mz^2-\delta\mw^2/\mw^2\right)$,
and $\delta e=e_0-e$, $\delta m_{W,Z}^2=m_{W,Z~0}^2-m_{W,Z}^2$
are the electric charge and gauge boson mass counterterms.
The $\delta g_Z$ factor is, in the BFG, an UV finite correction.
The $\Delta r$ parameter and the counterterms can be evaluated in perturbation theory
and we keep terms of \oa and \oaas \cite{Kniehl:1989yc,Degrassi:2003rw}.
For consistency, they have to be expressed in terms of $(G_\mu,\mw,\mz)$.
In addition to the redefinition of the overall weak coupling,
a second renormalization correction modifies the vector coupling $v_q$ of the $Z$ boson to the quarks:
$v_q = T_3^{(q)}-2 Q_q s^2 \to T_3^{(q)} - 2 Q_q (s^2+\delta s^2 +(c\,s/2)\, \delta Z_{AZ})$,
with $\delta Z_{AZ}$ the renormalization constant of the $\gamma-Z$ mixing.
In the BFG also this shift of $v_q$ is UV finite.
If we instead choose to relate $(g,g',v)$ to the $(\alpha,\mw,\mz)$ set of inputs
(dubbed $\alpha(0)$-scheme),
the replacement of the overall coupling is
$g_0/c_0\, Z_{ZZ}^{1/2}\to \sqrt{4\pi\alpha}/(s c) \left(1+\frac12 \delta g_Z\right)$,
while the redefinition of the vector coupling remains the same as in the other scheme.
The $\alpha(0)$-scheme choice is historically \cite{Sirlin:1980nh,Denner:1991kt}
one of the simplest EW renormalization input schemes,
but the low scale at which the fine structure constant is measured
yields in turn the appearance of large logarithmic corrections in the perturbative expansion.
The results depend on the value of the light-quark masses or, alternatively, on an experimental input
$\Delta\alpha_{had}(\mz)$ \cite{Jegerlehner:2001wq},
needed to evaluate the hadronic contribution to the running of the electromagnetic coupling at low scales.
The $G_\mu$-scheme is the most commonly adopted at hadron colliders,
because it reabsorbs in its definition large logarithmic corrections
and does not depend on the value of the light quark masses.
We take both as two extreme input possibilities and
we use the difference between the corresponding predictions
as a conservative estimate of the size of the missing EW higher order effects.
We remark that the size of the NLO-EW correction in the $G_\mu$ scheme is smaller than in the
$\alpha(0)$ case
because of an accidental partial cancellation between $\delta g_Z$ and $\Delta r$.
This fact is welcome in view of the phenomenological studies,
but it should be taken with care when estimating the size of the residual theoretical uncertainties.

The prediction of the hadron-level cross section requires to reabsorb the initial state singularities
due to collinear parton radiation in the definition of the physical proton PDFs.
Starting from NLO-EW, we need to subtract also the QED initial state collinear singularities,
to reabsorb them in the proton PDFs and to evolve the latter with DGLAP equations
that include a QED kernel.
The presence of a photon density in the proton yields new partonic scattering processes.
The hadron-level cross section is thus given by an extended set of processes,
with respect to those appearing in the QCD higher-order corrections.
The subtraction kernels at \oaas are based on the splitting functions
computed in Ref.\cite{deFlorian:2015ujt},
and allow to cancel from the partonic cross sections all the initial-state collinear singularities.
Two recent sets of proton PDFs accounting for NLO-EW corrections in the analysis
and evolved with a DGLAP QED kernel have been discussed in Refs.\cite{Ball:2017nwa,Harland-Lang:2019pla}.
These parameterisations have been obtained, starting from the same data set,
with two distinct modeling hypotheses, QCD only or full QCD-QED interaction,
and allow a meaningful comparison of the corresponding results.
We remark in fact that we have only two consistent options
to compute the best prediction for the total cross section:
$i)$ including only QCD radiative corrections and using PDFs evolved with only-QCD DGLAP kernels;
$ii)$ including QCD and EW corrections convoluted with QCD-QED PDFs. 
Given the impact of the EW corrections,
we consider the second option the only alternative to compute the best prediction for this cross section.

%%%%%%%%%%%%%%%%%%%%%%%%%%%%%%%%%%%%%%%%%%%

\section{Computational details}

In this letter, we complete the studies presented in Refs.\cite{Bonciani:2016wya,Bonciani:2019nuy},
with the analytical evaluation of all the \oaas contributions to the inclusive on shell $Z$ production cross section in terms of polylogarithmic and elliptic functions.
Since we are interested in the total $Z$-production cross section, our final result at parton level depends only on the variable $z$. 
We have to deal with two-loop virtual integrals in the double-virtual corrections, with two-body phase-space and 
one virtual loop integrals in the real-virtual case, and with three-body phase-space integrals for the double-real processes. 
In the last two cases, we use the reverse unitarity technique \cite{Anastasiou:2002yz,Anastasiou:2012kq} to transform the phase-space integrals 
into loop integrals satisfying the additional constraint imposed by on-shell-ness of the final-state particles. 
In this way, we can reduce all the integrals of all the partonic processes to the MIs via 
the integration-by-parts reduction technique \cite{Tkachov:1981wb,Chetyrkin:1981qh,Laporta:2001dd,Gehrmann:1999as}. 
In this phase, we exploit different computer codes: \texttt{Kira} \cite{Maierhoefer:2017hyi}, \texttt{LiteRed} \cite{Lee:2012cn,Lee:2013mka}, 
and \texttt{Reduze 2} \cite{Studerus:2009ye,vonManteuffel:2012np}. The resulting MIs are then computed solving the relevant 
differential equations \cite{Kotikov:1990kg,Remiddi:1997ny,Gehrmann:1999as,Argeri:2007up,Henn:2014qga,Ablinger:2015tua,Ablinger:2018zwz}. 
The massless MIs are the same as presented in Ref.\cite{Anastasiou:2002yz}. 
The two-loop virtual integrals with an off-shell $Z$ boson and internal massive line/s are 
studied in Refs.\cite{Fleischer:1997bw,Fleischer:1998nb,Aglietti:2003yc,Aglietti:2004tq,Aglietti:2004ki,Aglietti:2007as,Bonciani:2010ms,Kotikov:2007vr}. 
These off-shell integrals and their on-shell limits have been recomputed and checked 
using {\texttt{FIESTA}} \cite{Smirnov:2008py,Smirnov:2009pb,Smirnov:2015mct}. The corresponding results for the real-virtual and for the double-real 
cases were not available in the literature and have been computed expressly for this study.

The virtual contributions, proportional to $\delta(1-z)$, are constants found from the on-shell limit of the virtual MIs. 
They are computed by evaluating 
the corresponding generalized harmonic polylogarithms (GPLs) \cite{Goncharov:polylog,Goncharov2007,Remiddi:1999ew,Vollinga:2004sn}. 
The expressions of the real-virtual and double-real contributions are given in terms of $\delta(1-z)$, plus distributions and regular functions of $z$ 
which are expressed in terms of GPLs or cyclotomic Harmonic Polylogarithms \cite{Ablinger:2011te}.
We have used the packages 
{\texttt{GiNaC}} \cite{Vollinga:2004sn, Bauer:2000cp},
{\texttt{handyG}} \cite{Naterop:2019xaf} and
{\texttt{HarmonicSums}} \cite{Ablinger:2010kw,Ablinger:2013cf,Ablinger:2014rba},
for manipulation and numerical evaluation of all these polylogarithmic functions at various stages of our computation.
We particularly note that in double-real corrections, a set of three MIs arises, whose solution is given in terms of elliptic functions and
whose homogeneous behaviour has already been studied in Ref.~\cite{Aglietti:2007as}. 
We have obtained their complete solution with series expansions around $z=0$ and $z=1$ 
(see for instance \cite{Pozzorini:2005ff,Aglietti:2007as,Blumlein:2017dxp,Lee:2017qql,Lee:2018ojn,Bonciani:2018uvv,Blumlein:2019oas}).

The calculation of the MIs that depend on two different masses ($m_Z$ and $m_W$) is done performing an expansion of the integrand 
in powers of the ratio $\delta_{m^2}=(\mzsq-\mwsq)/\mzsq$. The coefficients of such expansion can be expressed, therefore, 
as a combination of the equal-mass MIs. While for the double-virtual corrections the expansion in $\delta_{m^2}$ is not strictly necessary, 
since the knowledge of the MIs for off-shell $Z$ boson allow for an exact calculation with $m_W \neq m_Z$, 
the reduction of one mass scale in the computation of the real emission processes
reduces effectively the complication of the calculation. 
The results, presented in this letter, are obtained using the expansion up to second order in $\delta_{m^2}$, wherever needed.
We stress that the couplings of the $Z$ boson to fermions are expressed in terms of the physical value of the weak mixing angle $\sin^2\theta_W=1-\mwsq/\mzsq$.

%%%%%%%%%%%%%%%%%%%%%%%%%%%%%%%%%%%%%%%%%%%%%%%

\section{Results}

In this section we present the numerical results for the inclusive 
total cross section for the production of an on-shell $Z$ boson in 
proton-proton collisions at the LHC. They are computed using the 
following values of the input parameters:
$\sqrt{S}=13\,{\rm TeV}$,
$m_H=125.0\, {\rm GeV}$,
$\mw=80.358\, {\rm GeV}$, 
$\mz=91.153\, {\rm GeV}$, 
$m_t=173.2\, {\rm GeV}$, 
$\alpha^{-1}=137.035999074$,
$G_\mu=1.1663781\,10^{-5}\, {\rm GeV}^{-2}$
and $\Delta\alpha_{had}(\mz)=0.027572$,
where $m_t$ and $m_H$ are the top quark and Higgs boson masses.
To present the results, we arrange $\sigma_{tot}$
as follows:
\begin{equation}
 \sigma_{tot} = \sigma_{LO} + \sigma_{10} + \sigma_{01} + \sigma_{11} + \sigma_{20}
\end{equation}
where $\sigma_{ij}$ indicates the sole contribution from the relative perturbative order ${\cal O}(\alpha_s^i \alpha^j)$
with respect to the Born.
We define the combinations
\begin{align}
 B_1 &= \sigma_{LO} + \sigma_{10} + \sigma_{20} \,,
 \\
 B_2 &= \sigma_{LO} + \sigma_{10} + \sigma_{01} + \sigma_{20}  \,,
 \\
 B_3 &= \sigma_{LO} + \sigma_{10} + \sigma_{01} + \sigma_{11} + \sigma_{20}
 \\
 B_3^{\gamma} &= \sigma_{LO} + \sigma_{10} + \sigma_{01} + \sigma_{11}^{\gamma} + \sigma_{20}
\end{align}
to study the effect of individual contributions. 
We note that $\sigma_{11}^{\gamma}$ denotes the mixed NNLO-QCD$\times$QED corrections 
whereas $\sigma_{11}$ denotes the full QCD-EW set.
All the $B_i$s are computed with the {\tt NNPDF31\_nnlo\_as\_0118\_luxqed\_nf\_4} \cite{Ball:2017nwa} proton PDF set,
which is evolved with both QCD and QED kernels.
We also define $A_1$ as the quantity corresponding to $B_1$ but evaluated with 
the {\tt NNPDF31\_nnlo\_as\_0118\_nf\_4} PDF set, the version with only-QCD DGLAP evolution.
We use both the {\tt LHAPDF-6} \cite{Buckley:2014ana} and {\tt HOPPET} \cite{Salam:2008qg} to interface the PDFs. 
The relevant value of the strong coupling constant is $\alpha_s(\mz)=0.1127$.

%%%%%%%%%%%%%%%%%%%%%%%%%%%%%%%%%%%%%%%%%%%%%%%%%%%%%%%%%%%%%%%%%%%%%%%%%%%%%%%%%%%%%%%%%%%%%%%

In Table \ref{tab:schemeuncertainty}, we show the results at different perturbative orders
for the cross section, in the $G_\mu$ and $\alpha(0)$ schemes.
$A_{1}$ is only LO from the EW point of view and, as a consequence, the two predictions differ by 3.53\%.
The inclusion in $B_2$ of the NLO-EW corrections reduces the spread to 0.88\%.
The complete NNLO-QCD$\times$EW corrections in $B_3$ further reduce this uncertainty down to the 0.23\% level.
\begin{table}[h]%
 \begin{ruledtabular}
 \begin{tabular}{lccc}
order  & $G_\mu$ & $\alpha(0)$ & $\delta_{G_\mu-\alpha(0)}$ (\%) \\ \hline
   $A_1$          & 55787   & 53884    & 3.53  \\ \hline
   $B_1$          & 55651   & 53753    & 3.53  \\ \hline
   $B_2$          & 55501   & 55015    & 0.88  \\ \hline
   $B_3^{\gamma}$ & 55516   & 55029    & 0.88  \\ \hline
   $B_3$          & 55469   & 55340    & 0.23  \\
 \end{tabular}
 \end{ruledtabular}
   \caption{\label{tab:schemeuncertainty}
     Comparison of the cross sections (expressed in pb) computed at different orders, in the $G_\mu$ and
     $\alpha(0)$ input schemes setting the factorization and renormalization scales as $\mu_R=\mu_F=m_Z$.}
\end{table}
%%%%%%%%%%%%%%%%%%%%%%%%%%%%%%%%%%%%%%%%%%%%%%%%%%%%%%%%%%%%%%%%%%%%%%%%%%%%%%%%%%%%%%%%%%%%%%%
The comparison in the $G_\mu$-scheme between the QCD-only prediction $A_1$ and the 
best QCD-EW predictions $B_{3} (B_{3}^{\gamma})$ shows that the latter reduce the value of the cross section
by $-0.57\% (-0.49 \%)$.
For the sake of a technical comment,
the comparison between $A_1$ and $B_1$ gives an estimate of the impact of DGLAP joint QCD and QED
evolution on the quark and gluon densities, yielding a reduction of the cross section by $-0.24\%$.
We stress that $B_1$ can not be considered a physical prediction,
because it lacks the EW corrections and the photon-induced partonic processes.
The size of $\sigma_{11}^{\gamma}$ is 0.03\% of the Born, 
while $\sigma_{11}$ is negative and 
larger than $\sigma_{11}^{\gamma}$ by almost a factor of three, as also observed in \cite{Buccioni:2020cfi}.
%%%%%%%%%%%%%%%%%%%%%%%%%%%%%%%%%%%%%%%%%%%%%%%%%%%%%%%%%%%%%%%%%%%%%%%%%%%%%%%%%%%%%%%%%%%%%%%
% 
\begin{table}[h]
 \begin{tabular}{l c c c}
   \hline\hline
   ~~ ~~~~ &  $G_{\mu}$-scheme     & ~~ & $\alpha(0)$-scheme  \\ \hline
   $A_1$        & $55787_{-0.99\%}^{+0.26\%}$ &    & $53884_{-0.99\%}^{+0.26\%}$ \\ \hline
   $B_2$        & $55501_{-0.99\%}^{+0.26\%}$ &    & $55015_{-1.26\%}^{+0.52\%}$ \\ \hline
   $B_3$        & $55469_{-1.01\%}^{+0.28\%}$ &    & $55340_{-1.13\%}^{+0.37\%}$ \\ \hline\hline
 \end{tabular}
   \caption{\label{tab:mufuncertainty}
     Dependence of the cross sections,
     expressed in pb and computed at different perturbative orders,
     under variation of the factorization scale $\muf$,
     keeping $\mu_R$=$m_Z$.}
\end{table}
%%%%%%%%%%%%%%%%%%%%%%%%%%%%%%%%%%%%%%%%%%%%%%%%%%%%%%%%%%%%%%%%%%%%%%%%%%%%%%%%%%%%%%%%%%%%%%%

In Table \ref{tab:mufuncertainty} we show the results obtained under variation of the factorization scale $\muf=\xi_F\mz$ in 
the range given by $\xi_F=(\frac12,1,2)$, keeping $\mu_R$=$m_Z$. Since the NLO-EW corrections present in $B_2$ are only LO from 
the QCD point of view, they vary accordingly with $\muf$, while the inclusion of the NNLO-QCD$\times$EW terms in $B_3$ stabilizes the results. 
The improvement from $B_2$ to $B_3$ is more evident in the $\alpha(0)$-scheme, where the size of the NLO-EW corrections is larger than in 
the $G_\mu$-scheme. The $A_1$ prediction has a variation by $+0.26\%,-0.99\%$, which increases in the $B_2$ case to $+0.52\%,-1.26\%$. 
The improvement induced by the mixed QCD$\times$EW corrections in $B_3$ brings the uncertainty down to $+0.37\%,-1.13\%$.

In Table \ref{tab:muruncertainty} we show the results obtained under variation of the renormalization scale  $\mur=\xi_R\mz$ in the range 
given by $\xi_R=(\frac12,1,2)$, keeping $\mu_F$=$m_Z$.
\begin{table}[h]
 \begin{tabular}{l c c c}
   \hline\hline
   ~~ ~~~~ &  $G_{\mu}$-scheme     & ~~ & $\alpha(0)$-scheme  \\ \hline
   $A_1$        & $55787_{-0.13\%}^{-0.15\%}$ &    & $53884_{-0.13\%}^{-0.15\%}$ \\ \hline
   $B_2$        & $55501_{-0.12\%}^{-0.15\%}$ &    & $55015_{-0.12\%}^{-0.15\%}$ \\ \hline
   $B_3$        & $55469_{-0.13\%}^{-0.15\%}$ &    & $55340_{-0.05\%}^{-0.21\%}$ \\ \hline\hline
 \end{tabular}
   \caption{\label{tab:muruncertainty}
     Dependence of the cross sections,
     expressed in pb and computed at different perturbative orders,
     under variation of the renormalization scale $\mur$,
     keeping $\mu_F$=$m_Z$.}
\end{table}
%%%%%%%%%%%%%%%%%%%%%%%%%%%%%%%%%%%%%%%%%%%%%%%%%%%%%%%%%%%%%%%%%%%%%%%%%%%%%%%%%%%%%%%%%%%%%%%
% 
In both cases, $\xi_R=2$ and $\xi_R=1/2$ there is a mild reduction of the cross section with respect to the central $\xi_R=1$ value. 
This decrease is of almost -0.14\% in both $G_\mu$-scheme and $\alpha(0)$-scheme.

Evaluating the $\mu_F$ and $\mu_R$ dependence through the 7-point scale variation, we obtain our best prediction 
for $B_3$\footnote{PDFs uncertainties are not considered in the present analysis.}, as presented in Table \ref{tab:murfuncertainty}.
\begin{table}[h]
 \begin{tabular}{l c c c}
   \hline\hline
   ~~ ~~~~ &  $G_{\mu}$-scheme     & ~~ & $\alpha(0)$-scheme  \\ \hline
   $B_3$        & $55469_{-1.01\%}^{+0.65\%}$ &    & $55340_{-1.13\%}^{+0.68\%}$ \\ \hline\hline
 \end{tabular}
   \caption{\label{tab:murfuncertainty}
     Dependence of $B_3$, expressed in pb, under 7-point scale variation of the renormalization ($\mur$) 
     and factorization ($\muf$) scales.}
\end{table}
%%%%%%%%%%%%%%%%%%%%%%%%%%%%%%%%%%%%%%%%%%%%%%%%%%%%%%%%%%%%%%%%%%%%%%%%%%%%%%%%%%%%%%%%%%%%%%%
% 
% 

We have performed several checks for our computations. 
In particular, we have checked analytically and numerically the QCD$\times$QED part of our result against 
the one in Ref.\cite{deFlorian:2018wcj}, finding complete agreement. Moreover, we have checked numerically our inclusive 
cross section against the results presented in Ref.\cite{Buccioni:2020cfi}, finding agreement within the expected numerical accuracy.

In conclusion, we have presented the prediction of the on-shell $Z$ inclusive total production cross section, including the full 
set of \oaas corrections. The evaluation of the cross section requires the consistent usage of proton PDFs which are extracted 
and evolved including QED effects. The comparison with the NNLO-QCD predictions, based on only-QCD proton PDFs, shows a change of 
the central value by $-0.57\%$. The NNLO-QCD$\times$EW corrections reduce the impact of two sources of theoretical uncertainty, 
which were not previously under control: the input parameters and the factorization scale variation uncertainties. 
The increased precision of the prediction of the $Z$-production cross 
section may have an impact on the determination of the proton PDFs
and on the hadron collider luminosity studies.

%%%%%%%%%%%%%%%%%%%%%%%%%%%%%%%%%%%%%%%%%%%%%%%%%%%%%%%%%%%%%%%%%%%%%%%

\begin{acknowledgments}
{\bf Acknowledgments.}
We would like to thank F. Caola, M. Delto, P. K. Dhani, M. Jaquier, J. Lang, K. Melnikov, V. Ravindran and R. R\"ontsch for useful discussions. 
A.V. is supported by the Italian Ministero della Universit\`a e della Ricerca (grant PRIN2017)
and by the European Research Council under the European Unions Horizon 2020 research and innovation Programme (grant agreement number 740006).
R. B. is partly supported by the italian Ministero della Universit\`a e della Ricerca (MIUR) under grant PRIN 20172LNEEZ.
The research of F.B. was partially supported by the ERC Starting Grant 804394 {\sc{hip}QCD}. 
R.B. and N.R. acknowledge the COST (European Cooperation in Science and
Technology) Action CA16201 PARTICLEFACE for partial support.
\end{acknowledgments}

\bibliography{BBRV}

\end{document}